\def\year{2023}\relax
\documentclass[letterpaper]{article} 
\usepackage{aaai22}  
\usepackage{times}  
\usepackage{helvet}  
\usepackage{courier}  
\usepackage[hyphens]{url}  
\usepackage{graphicx} 
\usepackage{natbib}  
\usepackage{caption} 
\usepackage{subcaption}
\DeclareCaptionStyle{ruled}{labelfont=normalfont,labelsep=colon,strut=off} 
\usepackage{algorithm}
\usepackage{algorithmic}
\usepackage{balance}
\usepackage{newfloat}
\usepackage{listings}
\floatstyle{ruled}
\newfloat{listing}{tb}{lst}{}
\floatname{listing}{Listing}
\title{Unveiling the Dynamics of Censorship, COVID-19 Regulations, and Protest: \\ An Empirical Study of Chinese Subreddit r$/$china$\_$irl}
\author {Siyi Zhou, Luca Luceri, Emilio Ferrara}
\affiliations{
    Annenberg School of Communication and Journalism \& USC Information Sciences Institute\\
    University of Southern California\\
}

\begin{document}

\maketitle

\begin{abstract}
The COVID-19 pandemic has intensified numerous social issues that warrant academic investigation. Although information dissemination has been extensively studied, the silenced voices and censored content also merit attention due to their role in mobilizing social movements. In this paper, we provide empirical evidence to explore the relationships among COVID-19 regulations, censorship, and protest through a series of social incidents occurred in China during 2022. We analyze the similarities and differences between censored articles and discussions on r/china\_irl, the most popular Chinese-speaking subreddit, and scrutinize the temporal dynamics of government censorship activities and their impact on user engagement within the subreddit. Furthermore, we examine users' linguistic patterns under the influence of a censorship-driven environment. Our findings reveal patterns in topic recurrence, the complex interplay between censorship activities, user subscription, and collective commenting behavior, as well as potential linguistic adaptation strategies to circumvent censorship. These insights hold significant implications for researchers interested in understanding the survival mechanisms of marginalized groups within censored information ecosystems.

\end{abstract}

\section{Introduction}

\paragraph{The Path to the White Paper Movement.}

Since 2019, China has implemented strict COVID-19-related regulations in pursuit of a ``zero-COVID-19'' public health objective. However, in March 2022, an abrupt and rashly-planned lockdown in Shanghai led to shortages of essential supplies and widespread unemployment \cite{pr:2022}. Consequently, citizens turned to Chinese social media platforms, such as Weibo and WeChat, to express their dissatisfaction and frustration. Despite stringent censorship measures aimed at curtailing the spread of sensitive critiques, social media users persisted in voicing their opposition to the ``zero-COVID-19'' regulations. To circumvent censorship, users employed homonyms, shared screenshots of censored content, and edited film clips like \textit{Les Miserables} to portray their experiences. These collective actions, referred to as the \textit{Shanghai lockdown protest} or the \textit{414 cyberprotest}, set the stage for a series of protests that unfolded from October to December 2022.

On October 13, 2022, an image of a banner on Beijing's Sitong Bridge bearing a slogan opposing COVID-19 testing, lockdowns, censorship, and dictatorship went viral online but was quickly censored. Later that month, images of Foxconn workers striking and protesting against the lockdown in a bid to return home were also censored. In early November, videos of crowds assisting a man shouting ``Bu zi you, wu ning si'' (\textit{Give me liberty, or give me death}) while escaping the police were similarly censored. Despite censorship efforts, protesters continued to voice their dissent. For example, an article featuring mosaic text, designed to mimic censorship, was circulated by protesters and subsequently censored. In late November, citizens spontaneously gathered for a public memorial service for the Urumqi fire victims, brandishing white papers to mock censorship and lockdowns. These white papers, which were banned from sale during the movement, symbolized the year 2022---a year characterized by COVID-19-19 regulations, censorship, and protests in China (\textit{cf.} timeline of events in Table~\ref{Tab:Tcr}).

 \begin{table*}[t]\small
 \caption{Detailed timeline of the \textit{Shanghai lockdown protest}.}
\begin{tabular}{@{}llp{0.5\linewidth}@{}}
\hline
Date               & Incident                       & Description \\ \hline
April 2022          & Shanghai Lockdown              & A sudden and lengthy lockdown in Shanghai displeased citizens with economic effects and generated outrage on social media sites. The complaints were difficult to suppress despite of the censorship. \\
October 13, 2022  & Beijing Sitong Bridge Banner   & A banner with anti-lockdown and pro-democracy banner was hung on the parapet of Sitong bridge in Beijing on the eve of the 20th National Congress of the Chinese Communist Party.                      \\
November 22, 2022 & Foxconn Strike                 & Workers of the Foxconn were locked in the factory in response to the zero-COVID-19 policy while ensuring to keep the facetory running. Videos of the workers scaled through barriers and fled home with bare foot went viral online.\\ 
November 24, 2022  &  Chongqing Superman    & A man from Chongqing was filmed giving an anti-lockdown speech, shouting ``Give me freedom, or give me death.'' The crowd helped him escaped from police even though he was arrested eventually.                     \\
November 24, 2022 & Urumqi fire                    & A fire in a Urumqi building caused 10 deaths. The victims were locked at home during the disaster because of the ZERO-COVID-19 policy.\\
November 26, 2022 and onwards & White Paper Movement  & Chinese citizens located different cities and Universities, individuals located all over the world, voluntarily attended memorial meetings for the Urumqi fire victim. During the public gathering, they shouted the slogans from the previous incidents and held up blank white paper to protest against COVID-19 regulations and censorship. \\\hline
\end{tabular}
  \label{Tab:Tcr}
\end{table*}

\paragraph{Theoretical Background.}

For decades, scholars have delved into the mechanisms that drive the mobilization and organization of social movements. The concept of organization, which underlies the unity of social movements, has been a subject of continuous inquiry due to its significant influence on the ability of social movements to challenge established power structures \cite{c:13}. While some theorists consider formal organization essential for preventing disintegration and primitivism \cite{h:59}, others argue that excessive organizational structure can undermine the disruptive potential and hinder social activism \cite{mpp:15, pc:77}. Tarrow posits that an optimal degree of organization is necessary to sustain momentum and achieve a social movement's goals while maintaining a structure flexible enough to capitalize on informal networks \cite{t:98}.

The emergence of the digital communication era has facilitated new forms of collective action through information sharing and user connectivity on online social media platforms. These features have amplified the frequency and impact of social movements \cite{m:94,g:06,b:03}. Online social media  introduced a novel domain of organizational structure and information dissemination for collective action, consequently drawing significant scholarly interest. Most contemporary studies on social movements, such as the Occupy Wall Street and Gezi Park movements, have centered on the dissemination of information during and after protests \cite{cfmf:13, cfmmf:13, varol2014evolution}.

The convergence of digital and physical spaces has generated new opportunities for collective action. This hybrid space functions as a platform for individuals to forge symbolic communities and foster transformative practices \cite{c:13, bs:12}. While physical protests rely on public demonstrations of solidarity, such as chanting slogans and displaying signs, cyberspace enables more personalized expressions of emotions and individual motivations \cite{allen2021pictures, chang2022justiceforgeorgefloyd}.

In 2022, a series of protests against COVID-19 regulations and censorship emerged in both physical and digital spaces in China. The enforcement of censorship has rendered overt protests and social movements seemingly unattainable, particularly under the ``zero-COVID-19'' regulation and lockdown policy. Online content censorship has obstructed people's ability to share information about social incidents, while stringent lockdown policies have impeded offline connection. The 2022 protests in China present a fertile area for researchers to examine the formation of unity and community amidst disintegration and surveillance, as well as to investigate the actions of affected, vulnerable populations in such constrained environments.

\subsection{Contributions of this Paper}

In this paper, we present an empirical study on censored articles, government censorship activity, and user online behavior within a Chinese-speaking Reddit community during the 2022 COVID-19 protests in China. This paper aims to glean actionable insights into the dynamics of COVID-19 regulation, censorship, and protest through an exploratory analysis that addresses the following research questions:

\begin{itemize}
\item[\textbf{RQ1}] \emph{How do the topics of discussion in r/china\_irl interact with the articles censored by the Chinese government?}
\item[\textbf{RQ2}] \emph{How do subscription and commenting behavior in r/china\_irl relate to government censorship behavior? Can the temporal dynamics predict one behavior based on another?}
\item[\textbf{RQ3}] \emph{Do individuals adapt and modify their language or linguistic style to evade censorship?}
\end{itemize}

To address these questions, we compile a heterogeneous dataset comprising censored articles, whose access is no longer guaranteed to the public via their original URLs, and one year of user activity in the most popular Chinese-speaking subreddit, r/china\_irl. By employing tokenization and topic modeling in the Chinese language (ensuring that newly invented Chinese phrases can be accurately detected and tokenized), we offer a quantitative and qualitative analysis of the unfolding protest and related online discussions. Our exploratory analysis uncovers novel social phenomena and poses several potential questions to initiate and stimulate further research.

\section{Related Work}
When studying social protest, multiple dimensions are considered, from mobilization to participant demographics. Technological advancement allows scholars to gain more insights into social protests and movements. Scholars first study protests through news articles  \cite{uu:11,soy:16,ls:17}. In studying the Tweeting behavior, Steinert-Threlkeld found that increased protests generate increased message coordination of specific hashtags \cite{smvf:15}.  Independent studies pointed out the important role of image in online protest mobilization in the case of Black Lives Matter \cite{cw:19, chang2022justiceforgeorgefloyd}.  MMCHIVED collects images from Twitter and identifies demographic information from the protesting images during the Venezuela and Chile protest \cite{sj:22}.  Most of the protest data covers images, hashtags, or news articles. Existing tools and studies help us understand the demographics and mobilizations of protests, yet limited information for us to investigate detailed topics and participant agendas of the protests. 

Existing studies on Chinese social protest mainly focus on the non-sensitive ones. The CASM database uses convolution neural networks and LSTM to identify possible offline protests on social media \cite{zp:19}. However, the collection of protest information from Chinese social media platforms is heavily filtered by non-sensitive events if collected afterward. In order to gain more insights into the highly sensitive social protests in China, researchers mostly rely on in-depth interviews,  court files, and newsletters overseas \cite{l:21}. Massive first-hand data is extremely limited.  In search for related content about the white paper movement or protests against COVID-19 regulations in Chinese social media and newspapers, almost nothing can be found about such information. The Wayback Machine, however, confirms the existence of those incidents as some of the web pages were stored. 

Scholars investigating censorship have focused on how governments use censorship tactics to suppress protests and how individuals collectively circumvent censorship. The 2019 Hong Kong protests provide an illuminating case study where censorship was identified as having the ability to both demobilize and rally the masses by selectively reporting threats and downplaying events in newspapers  \cite{mw:23}. Web-blocking and DoS attacks were identified as the primary censorship tactics used to repress protests in several countries, including Iran, Malaysia, Ukraine, and Venezuela \cite{kwd:23}. When faced with censorship, individuals often seek to circumvent it to obtain additional information, with crises such as COVID-19 acting as a catalyst for such behavior \cite{chrs:22}. The 2022 protests in China saw censorship initially used to repress protests against zero-COVID-19 regulation, but ultimately backfired and sparked the white paper movement that protested against censorship.

The case of the 2022 protests in China is not monocausal. A series of protests are caused by the lockdowns in Shanghai, the zero-COVID-19 regulation in Beijing, the secondary crisis in Urumqi, and the censorship of freedom of speech nationwide. Collective attention is thought to exhaust fast if the production and consumption gradient is high \cite{lmhl:19}. However, public attention on COVID-19 regulation and censorship was high and lasting. Limited resources help us understand how people are able to gather together when  being locked at home under censorship, and what are the connections between a series of protests, and post-protest reactions. Thus we collected data from censored articles and Chinese Reddit submissions and comments to help study the extraordinary year of events in China. 

\begin{figure*}[t]
\centering
\includegraphics[width=.97\textwidth]{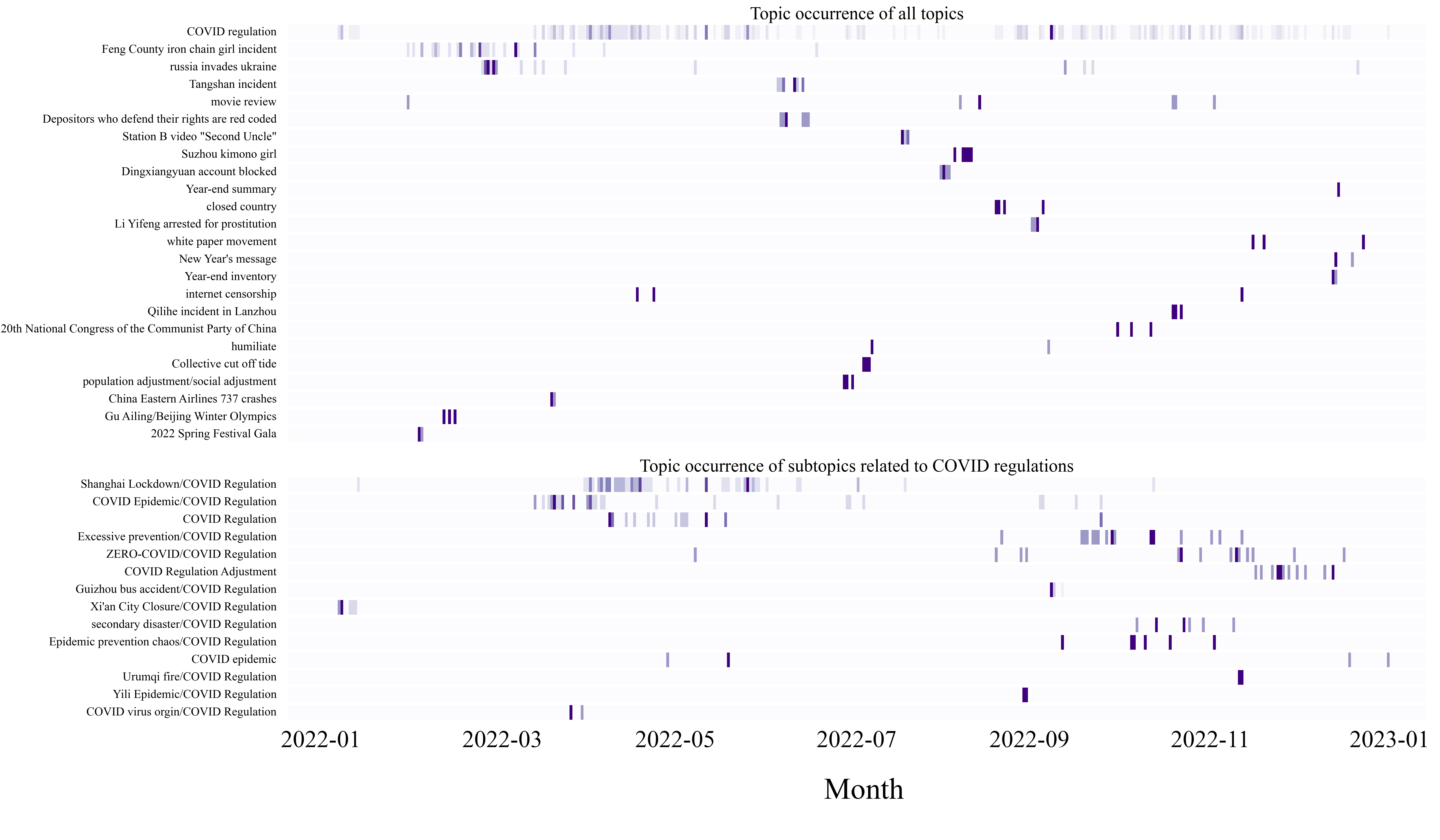} 
\caption{Event plot to show frequency distribution of when articles of a topic are censored. }
\label{fig2}
\end{figure*}

\begin{figure*}[t]
\centering
\includegraphics[width=.99\textwidth, clip=true, trim= 200 200 200 200]{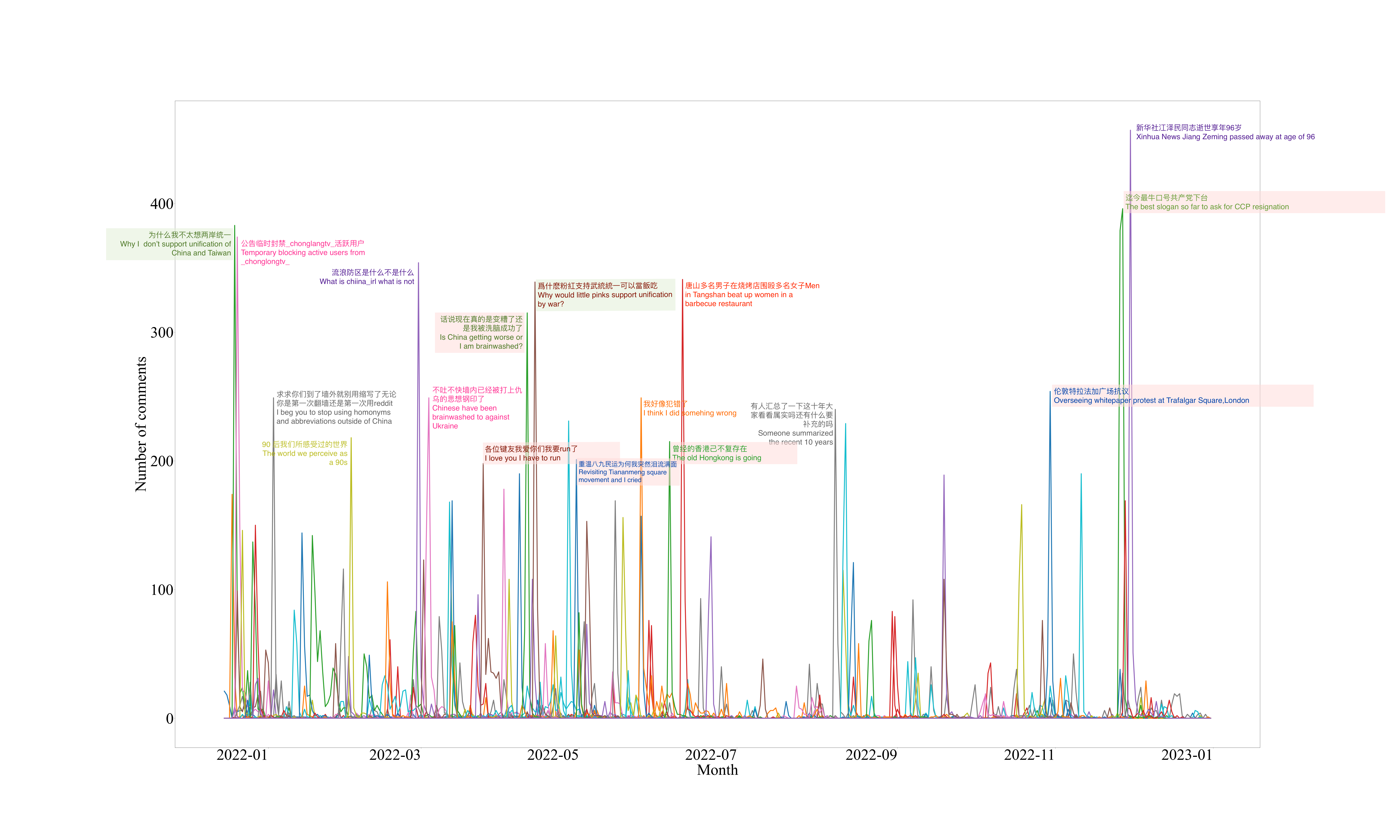} 
\caption{$r/china\_irl$ submission activity that describes the number of comments of each submission on each day. Submission with over 200 comments in one day is labeled with their topic.}
\label{fig1}
\end{figure*}

\section{Methodology}

This section elucidates the process of data collection, the reasoning behind it, and the methods of data processing.

\subsection{Data Collection}
The dataset presented in this paper is composed of two parts: (i) 1,249 censored articles published since 2019,  (ii) Metadata and 1.6 million comments from the Reddit community r$/$china$\_$irl shared from December 2021 to January 2022.

The censored articles are collected from \textit{China Digital Times} (CDT) and contain the title, content, original link, and archival dates retrieved from the \emph{Internet Archive}, also known as the \emph{Wayback Machine} (\url{https://archive.org/web/}). We further manually verified the original content and recorded all articles’ actual creation date by leveraging the Wayback Machine.  

We downloaded subscription statistics from \emph{Subreddit Stats} (\url{https://subredditstats.com/}) and all the comments of the most popular Chinese-speaking subreddit, r$/$china$\_$irl, from the Pushshift Reddit database (PRD) \cite{bzks:20}. The community-based style of the platform, in general, and the r$/$china$\_$irl subreddit, in particular, enables us to comprehend how people and ideas congregate in the online space. The subreddit r$/$china$\_$irl is created for conversations related to Chinese political and social issues after several Chinese forums, like \emph{Hupu kfq}, were shut down in 2019, and internet users' posts on Weibo were massively censored. While there exist other Chinese-speaking subreddits, r$/$china$\_$irl  represents the most prominent and widely used Reddit community with 176K active users and a rather neutral political stance.
With the aid of both the CDT and Reddit datasets, we anticipate observing some degree of correspondence between the censored articles and the topics discussed in the subreddit.

\subsection{Data processing}
Unlike many phonetic languages, Chinese is an ideogram language with a unique grammar system that renders the automated identification of phrases and stopwords a complex task. For example, one character can have multiple layers of meaning. This characteristic challenges computational analysis of Chinese textual content in many ways, including topic modeling, sentiment analysis, and sarcasm detection.

To overcome these hurdles in our analysis, we preprocess the Reddit data using both \textit{Jieba} (\url{https://github.com/fxsjy/jieba}) and \textit{NLTK} \cite{loper2002nltk}. We utilize a \textit{Hidden Markov Model} (HMM) and the \textit{Viterbi algorithm} in our tokenization process to detect newly invented Chinese words and phrases, especially homonyms, for different social incidents. This allows us to evaluate and compare the linguistic patterns, length of comments, and volume of submissions and comments at different times.
To understand the content and topics of the censored articles, we employ LDA to calculate word and topic frequency and BERTopic \cite{grootendorst2022bertopic} to classify the topics in the dataset. While there are many pre-trained topic models, most are trained on English or phonetic languages. 

To obtain the most accurate topic classifications, we label and train our own dataset. CDT provides the topic of the censored articles in the form of tags, which we further refine to better generalize the topics. For example, articles discussing the secondary crisis caused by the lockdown are tagged by CDT as either ``Shanghai Lockdown'', ``secondary crisis'',  or ``COVID-19 regulation''. We generalize such a group of articles with tangible social issues and policies, such as “Shanghai Lockdown/COVID-19 regulation” to avoid bias. This helps us determine the main topics of discussion and evaluate their relevance in our analysis.


To investigate the relationship between Reddit user behavior and government censorship behavior, we employ Granger causality and Pearson correlation analysis. However, the relationship between censorship, commenting, subscribing, and linguistic patterns are not straightforward. 
We use Granger causality tests to estimate the time lag for variable X to respond to variable Y or the number of time steps one action precedes another. Then, we test the correlations of each pair of the two time series with the suggested time shift provided by the Granger test.
It's worth noting that causality cannot be always ascertained, as this multivariate relationship forms a complex system where the variables could be inter-affecting each other. Consequently, for this study we limit ourselves to determining the correlations of the variables rather than their causal relationships.

\section{Results}

\subsection{RQ1: Topics Interplay Between Censored Articles and Reddit Discussions}

From the CDT dataset, the summary of when articles regarding a certain topic is published and censored shows us the most sensitive topics of different time periods in 2022. Previous assumption is that people will stop mentioning or discussing a sensitive topic after being censored for a period. However, Figure \ref{fig2} suggests that while topics that touched upon one social incident were censored for a concentrated period in 2022, the frequency of censoring activities against topics related to social issues, especially COVID-19 regulations, is more dispersive. Under the discussion of COVID-19 regulation topic, subtopics are generated in regards to multiple different social incidents, such as the Shanghai lockdown, Beijing Sitong Bridge banner, Urumqi fire, and White Paper protest, as shown in Figure \ref{fig1}. 

Articles integrating multiple social incidents are labeled with the social issues they attempt to discuss, such as Zero-COVID-19 and the COVID-19-pandemic as a whole. Those COVID-19-related articles, especially when discussing a social issue, tend to recur when a COVID-19-related incident happens regardless of the risk of censorship. Articles on such topics keep being published even though the author might already know the article has a high risk of being censored. This behavior contributes to the outrage of the White Paper movement later on.

The recurrence phenomenon of discussing high-risk censor-prone topics in China also appears in r$/$china$\_$irl. Figure \ref{fig1} graphs the number of comments each submission receives each day in 2022. The graph excludes 3 outlier submissions (moderated submissions peaking over 1000 comments in one day) in order to get a more general picture of the conversations. Submissions that received over 200 comments in one day are labeled with their titles. Titles related to the same topic are highlighted with the same color. Preliminary observation discovers that most of the submissions that receive over 200 comments in one day are not talking about the same topic. However, topics about the Ukraine war and protests in China are repeatedly mentioned in submissions and popularly discussed in comments. The topic of protest in China was repeated significantly more times than the Ukraine war topic. 

While the censored articles collected from Chinese social media and the subreddit forum share characteristics of topic recurrence, their emphasis are slightly different. The censored articles are more oriented towards examining one or more specific consequences of the COVID-19 regulation and explaining why the situation needs to be changed. However, the popular recurring Reddit submissions, although under the big picture of COVID-19 regulation, tend to repetitively bring up topics in regard to protest behavior. Instead of discussing why people are suffering, Reddit users compare COVID-19-related protests to other major protests in China, such as Tiananmen Square movement and Hong Kong Umbrella movement, as well as sharing overseas protests against China’s policies on COVID-19 and censorship abroad. 

Although paying attention to different aspects of a social issue,  the public agenda in r$/$china$\_$irl is influenced and partially set by the censored articles inside of China. In the next section, we look for other potential correlations between censorship activity and Reddit user behaviors in r$/$china$\_$irl.

\begin{figure*}[t]
 \centering
     \begin{subfigure}[b]{0.65\columnwidth}
         \centering
        \includegraphics[width=\columnwidth]{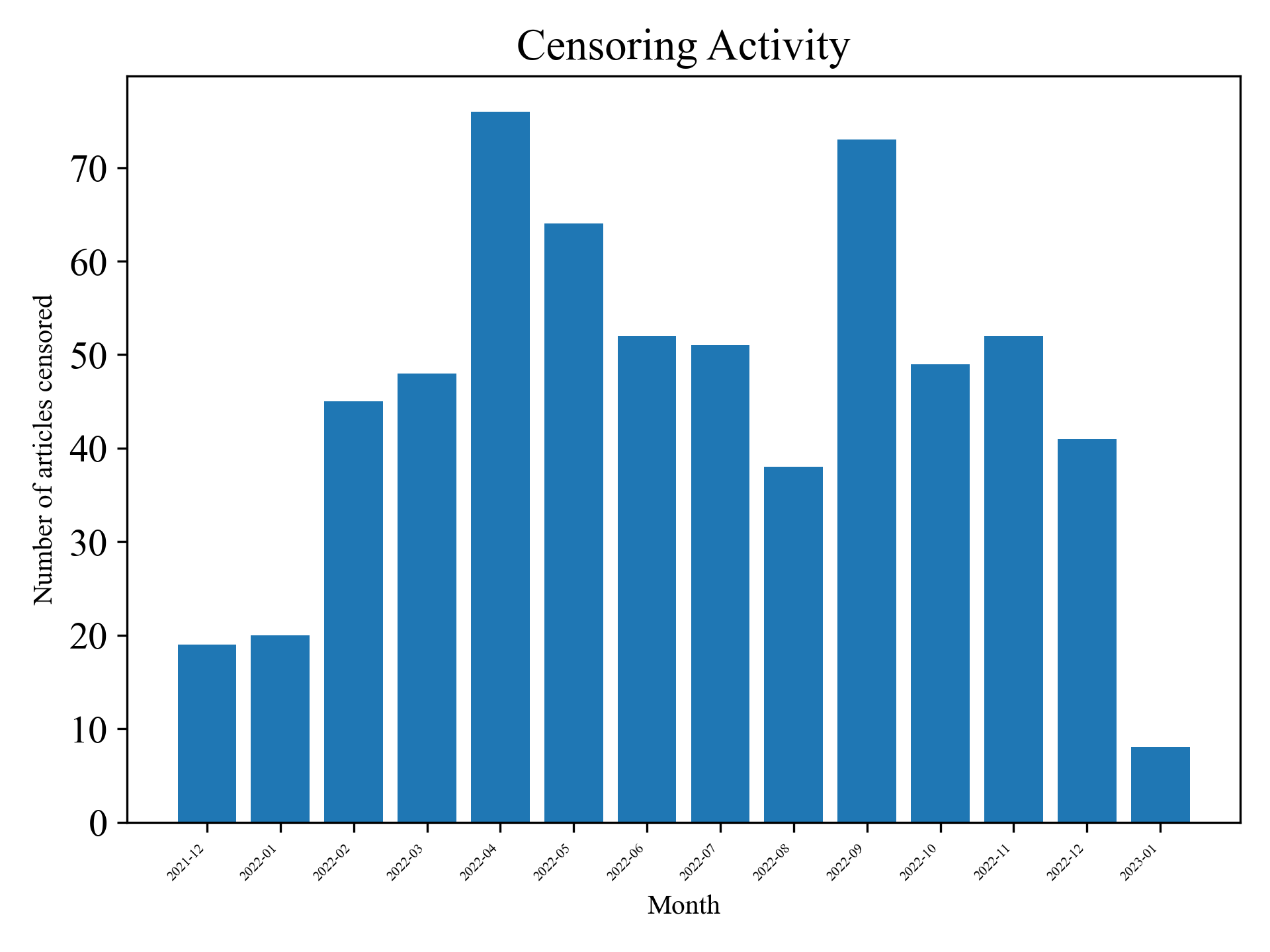} 
         \caption{Number of articles getting censored}
         \label{fig:y equals x}
     \end{subfigure}
     \begin{subfigure}[b]{0.75\columnwidth}
         \centering
         \includegraphics[width=\columnwidth]{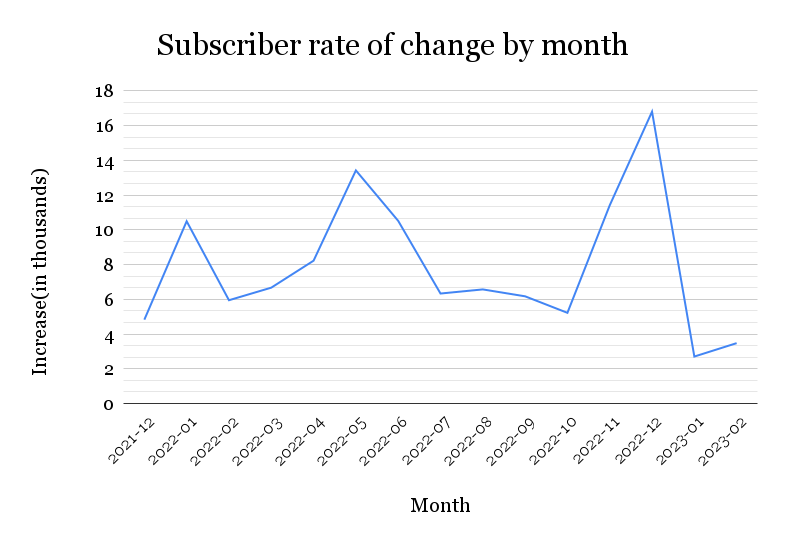} 
         \caption{Subscriber increase rate}
         \label{fig:three sin x}
     \end{subfigure}
     \begin{subfigure}[b]{0.65\columnwidth}
         \centering
         \includegraphics[width=\columnwidth]{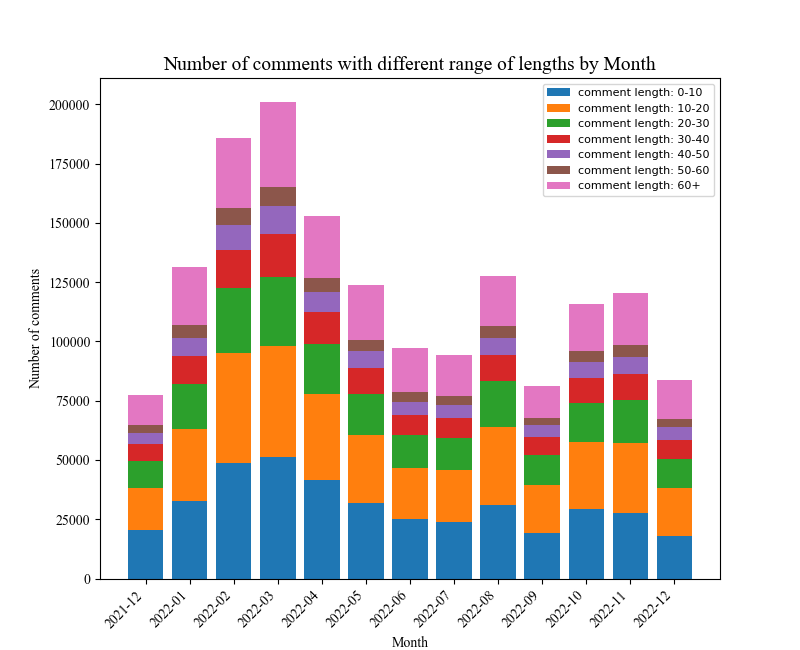}
         \caption{Number of comments}
         \label{fig:three sin x}
     \end{subfigure}

\caption{Monthly activities of censorship and Reddit: (a) Number of censored articles collected each month. (b) new subscribers to $r/china\_irl$ each month. (c) Stacked bar graph to summarize the number of comments generated each month. Each color represents the number of comments in the range of comment length specified in the legend. }
\label{fig3}
\end{figure*}

\subsection{RQ2: Temporal Dynamics and Interplay of Censoring Activity, User Subscriptions, and Collective Commenting}

Figure 3 presents the monthly activities of the number of articles getting censored, new subscriptions, and the number of comments on r$/$china$\_$irl. Although the graphs indicate a similar trend among the three activities, there are slight variations in the timing of their respective peaks.
To investigate the relationship between these variables, we first conducted a Granger causality test, as reported in Table 2. 

\begin{table}[h]\footnotesize
 \caption{Granger causality test}
\begin{tabular}{@{}l@{}c@{}c@{}rr@{}} 
\hline
Variable X & Variable Y  & Lag (w) & \ F-test &P-value \\ \hline
no. censored articles & no. comments &0&0&0 \\
no. comments& no. censored articles  &4&2.87&0.022 \\
no. censored articles & no. new subscribers &6&2.64&0.034 \\
no. new subscribers &no. censored articles  &1&8.80&0.0046 \\
no. new subscribers &no. comments &1&6.80&0.012 \\
no. comments&no. new subscriber &6&3.50&0.006 \\\hline
\end{tabular}
  \label{Tab:Tcr1}
\end{table}
\begin{table}[h]
 \caption{Censoring activity v.s. Commenting}
\begin{tabular}{@{}l@{}cr@{}}
\hline
Time Shift & Pearson corr. & p-value \\ \hline
No shift& 0.17&0.58 \\
Censorship 1 month after comments & 0.64 &0.026 \\
Censorship 2 months after comments& 0.50 &0.12 \\\hline
\end{tabular}
  \label{Tab:Tcr2}
\end{table}
\begin{figure}[t]
 \centering
     \begin{subfigure}[b]{0.3\columnwidth}
         \centering
        \includegraphics[width=\columnwidth]{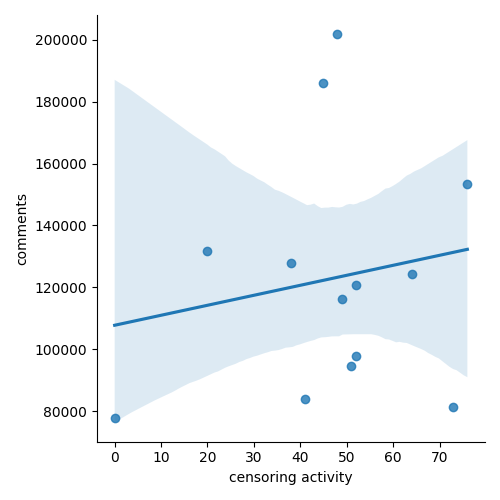} 
         \caption{No shift}
         \label{fig:y equals x}
     \end{subfigure}
     \hfill
     \begin{subfigure}[b]{0.3\columnwidth}
         \centering
         \includegraphics[width=\columnwidth]{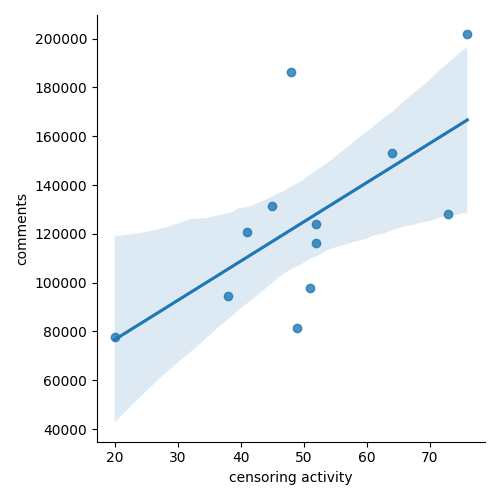} 
         \caption{Shift 1 month}
         \label{fig:three sin x}
     \end{subfigure}
     \hfill
     \begin{subfigure}[b]{0.3\columnwidth}
         \centering
         \includegraphics[width=\columnwidth]{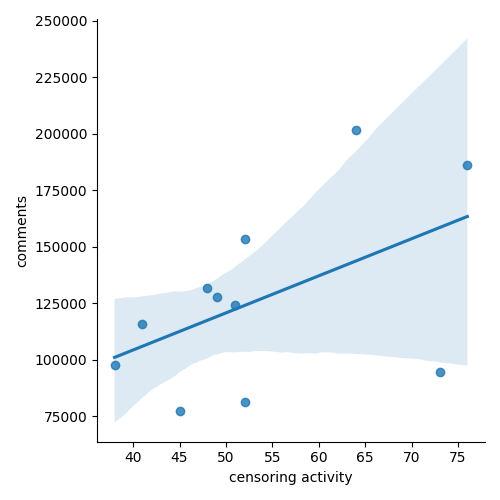}
         \caption{Shift 2 month}
         \label{fig:three sin x}
     \end{subfigure}
\caption{Correlation regressions of censorship activity v.s comments in the r$/$china$\_$irl}
\label{fig4}
\end{figure}

\begin{table}[h]\small
 \caption{Censoring activity v.s. Subscription}
\begin{tabular}{@{}l@{}cr@{}}
\hline
Time Shift & Pearson corr. & p-value \\ \hline
No shift& 0.38&0.22 \\
Censorship 1 month prior to subscription& 0.84 &0.009 \\
Censorship 2 months prior to subscription& 0.16 &0.64 \\\hline
\end{tabular}
  \label{Tab:Tcr3}
\end{table}
\begin{figure}[t]
 \centering
     \begin{subfigure}[b]{0.3\columnwidth}
         \centering
        \includegraphics[width=\columnwidth]{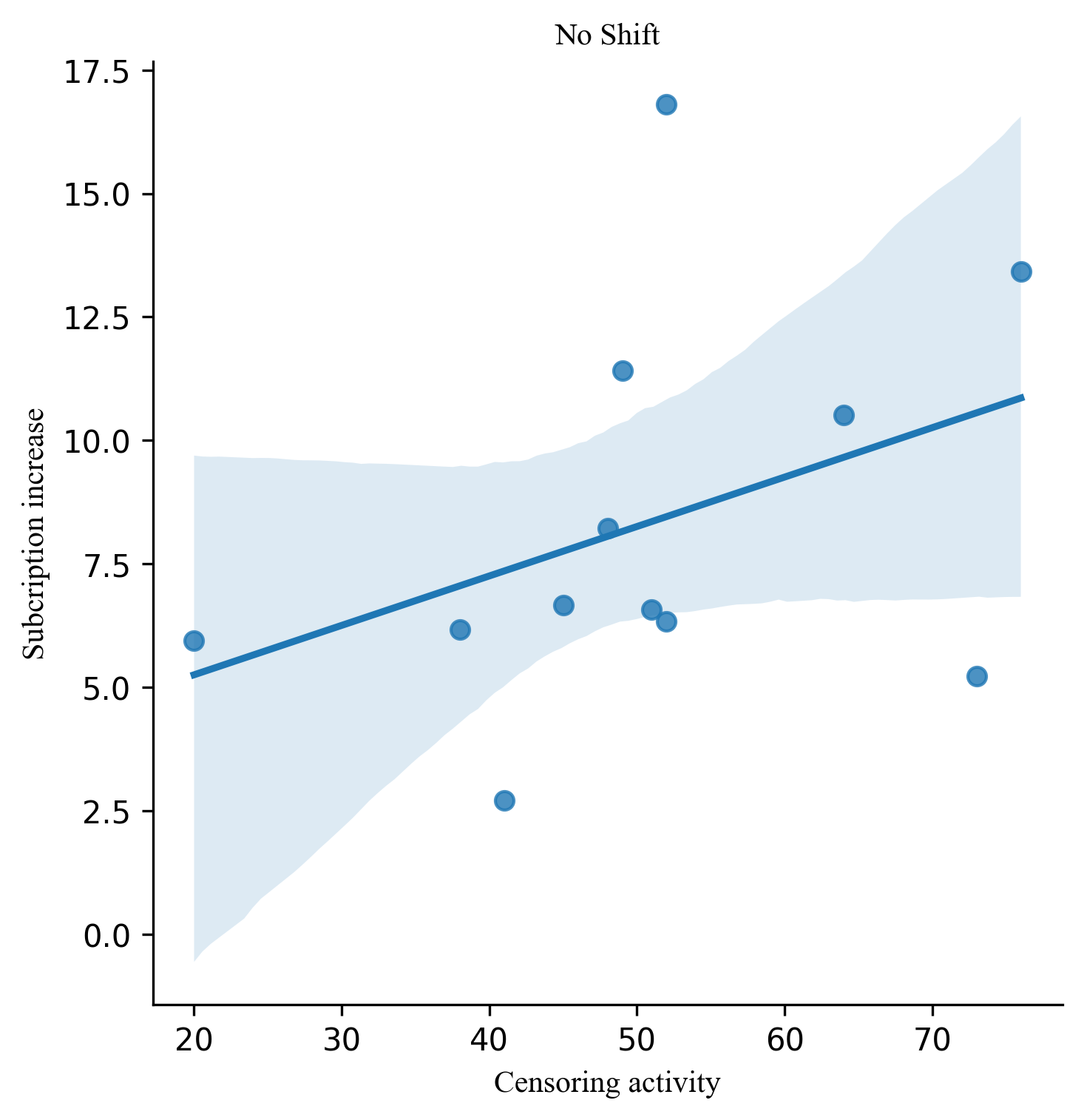} 
         \caption{No shift}
         \label{fig:y equals x}
     \end{subfigure}
     \hfill
     \begin{subfigure}[b]{0.3\columnwidth}
         \centering
         \includegraphics[width=\columnwidth]{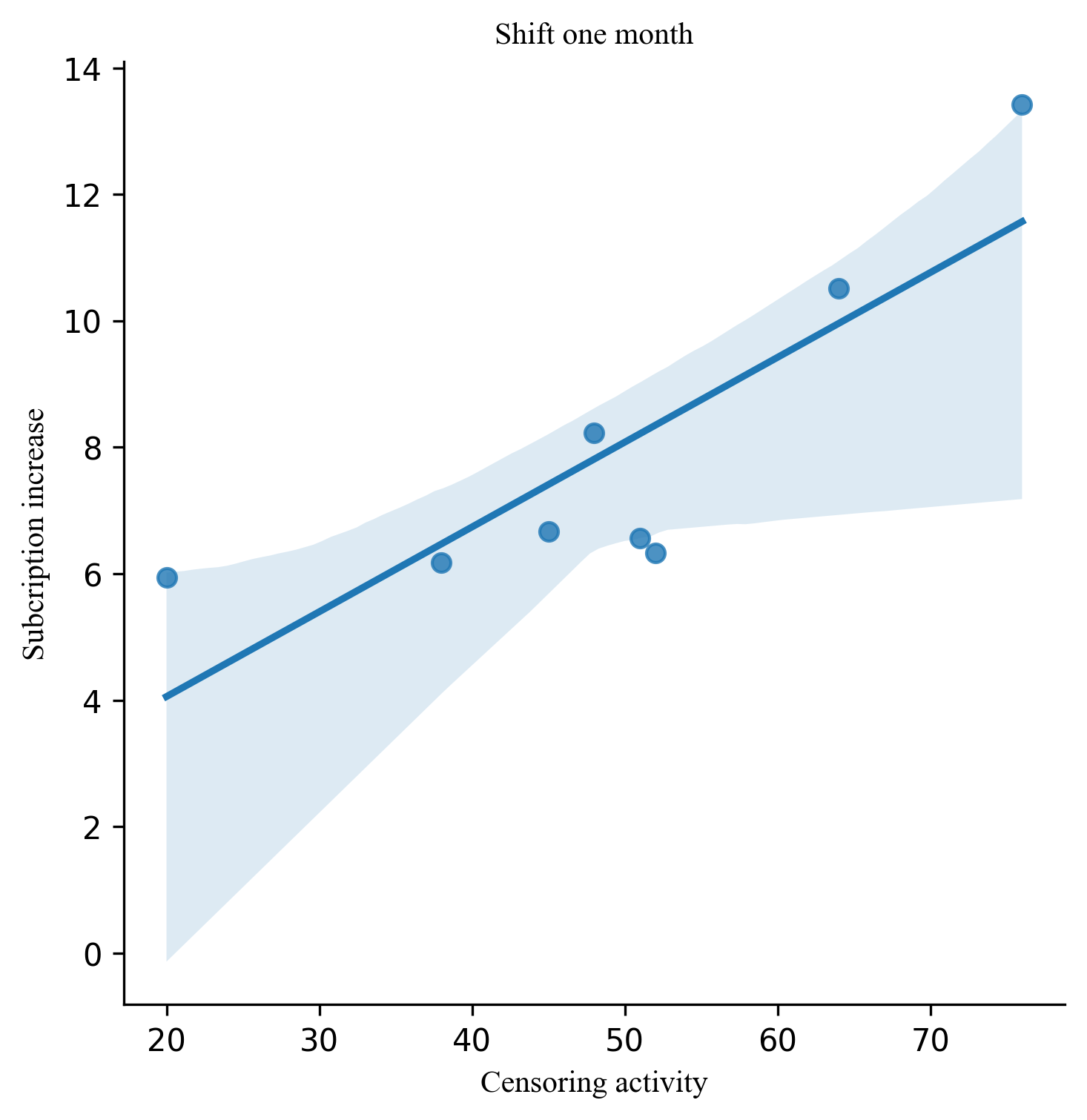} 
         \caption{Shift 1 month}
         \label{fig:Shift 1 month}
     \end{subfigure}
     \hfill
     \begin{subfigure}[b]{0.3\columnwidth}
         \centering
         \includegraphics[width=\columnwidth]{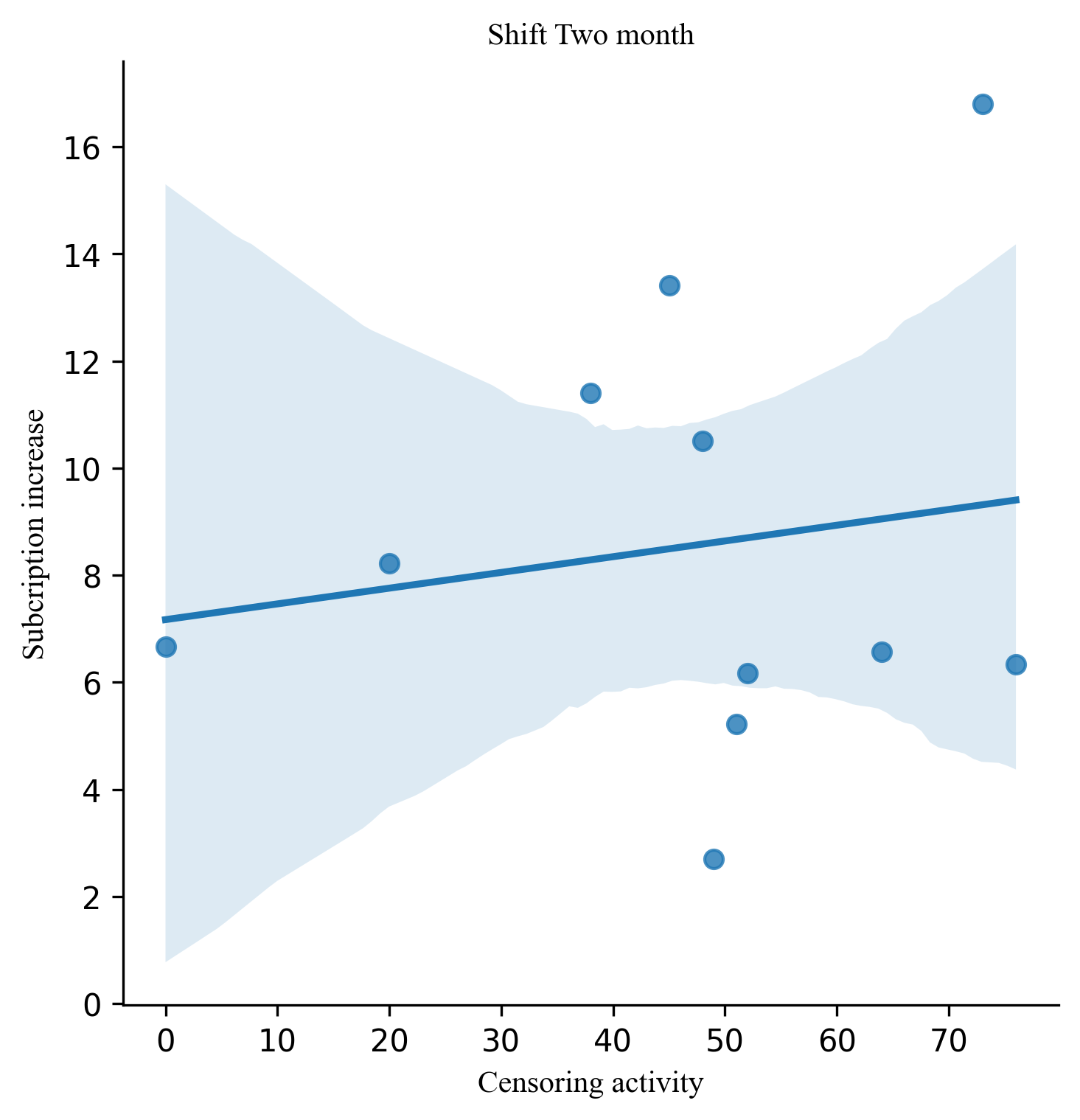}
         \caption{Shift 2 month}
         \label{fig:three sin x}
     \end{subfigure}

\caption{Correlation regressions of censorship activity v.s increase in subscribers in the r$/$china$\_$irl}
\label{fig5}
\end{figure}

Our results indicate that only the pair of censorship and comments pass the test, revealing that an increase in comments anticipates an increase in censorship activities with a lag of 4 weeks. The reverse relationship between comments and censorship is not statistically significant. The bi-directional relationships of the other pairs, subscriber and censorship, and comment and subscribers, fail the test, prompting a question about their interdependence. To further validate these relationships, we performed a Pearson correlation test using the suggested time lags.

To account for the impact of localized noise and focus on a broader, global scale analysis, we reduced the temporal resolution of our correlation analysis by measuring the correlations between variables at a coarser granularity. The regression models of censorship and comments, and censorship and subscribers are depicted in Figures \ref{fig4} and \ref{fig5}, respectively. The coefficient in Table \ref{Tab:Tcr2} aligns with the censorship/comment Granger test (coeff. = 0.64), indicating that an increase in censorship activity follows one month of the increase in comments. Table \ref{Tab:Tcr3} demonstrates a robust correlation (coeff. = 0.84) between censorship and new subscriptions, with the increase in subscriptions succeeding one month of the censorship activity. Although the subscriber/comment pair passes the Granger causality test in two directions, their p-values in the correlation test are too high to reject the null hypothesis, suggesting that the number of new subscribers and the number of comments are independent of each other.

The observed pattern wherein commenting behavior precedes censorship activities is unsurprising, as sensitive topics that attract high collective attention tend to raise government awareness and prompt censorship. \citet{chrs:22} highlights the spillover effect of the COVID-19 crisis, with increased downloads of Facebook and Twitter, as well as increased views of Wikipedia from users, geolocated in China, serving as a means of censorship circumvention. The increase in subscriptions following an increase in censored articles aligns with the common tendency of individuals to seek new outlets for expression when their original media platforms no longer allow them to speak freely.

\begin{figure*}[t]
\centering
\includegraphics[width=2.35\columnwidth]{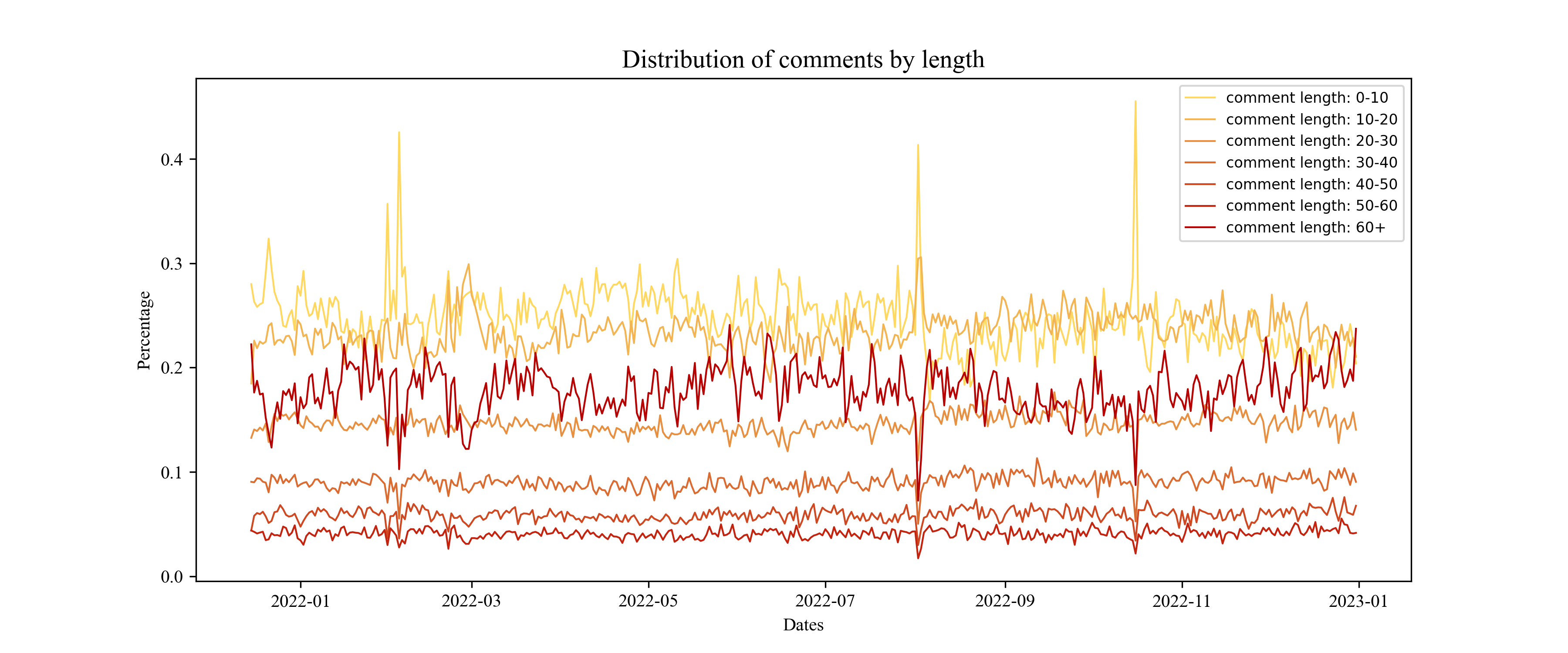} 
\caption{Proportion of different comment lengths of all comments each day.}
\label{fig6}
\end{figure*}

\subsection{RQ3: Linguistic Patterns}
To gain further insights into commenting behavior and linguistic patterns, we employed a grouping strategy that split comments by length. Figure \ref{fig6} revealed a significant proportion of short comments (ranging from 0-20 characters) and long comments (60+ characters), with moderate-length comments (20-60 characters) being less frequent. 

We further sought to perform content analysis to understand the conversations in the comments by tokenizing and removing stop words. However, we observed that 140,459 of the comments were transformed into empty strings after tokenization and the removal of stop words. Consequently, we retained all of the stop words and analyzed the entire textual content in Reddit comments. On average, 44.2$\%$ of all characters in comments were identified as stop words, with 10.2$\%$ of comments consisting solely of stop words. The proportion is substantial compared to the medium stop word frequency in text in English, which is typically around 1$\%$ as reported by a stop word distribution analysis in English with 19,449 articles using Gigaword corpus (\url{https://faculty.georgetown.edu/wilsong/IR/WD3.html}).

\begin{figure}[h!]
\includegraphics[width=\columnwidth]{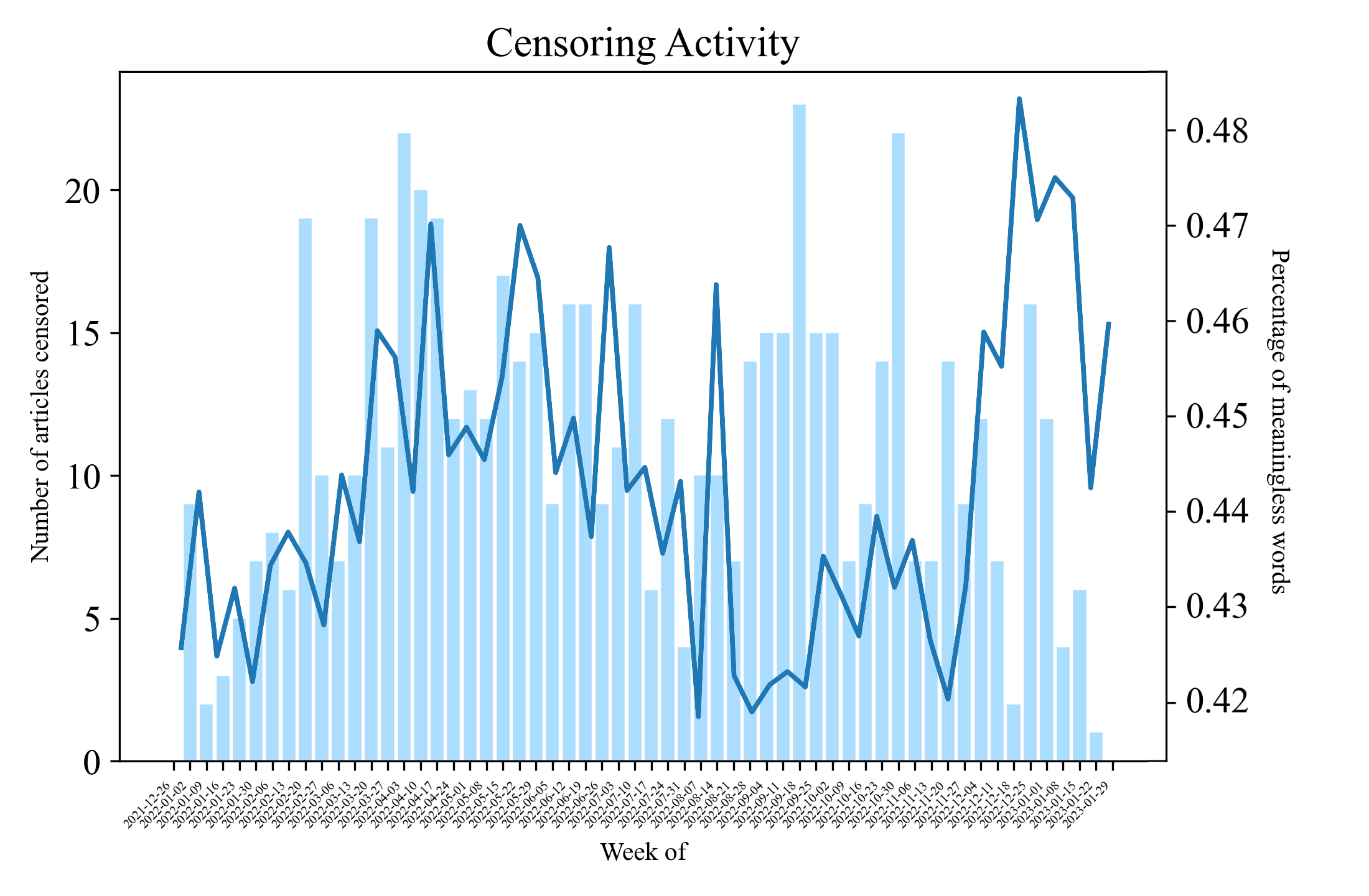} 
\caption{Stopwords usage frequency (line graph) and the censorship activity (bar graph) by week. }
\label{fig7}
\end{figure}

Interestingly, the frequency of stop-words usage bears a resemblance to the pattern of censorship activities in Figure \ref{fig7}, which led us to investigate whether censorship behavior had an impact on stop-word usage. Our Granger causality test showed that the increase in stop word usage follows the increase in censorship activities with a 3-week lag, whereas the converse failed the test. The correlation test further revealed a moderate correlation (coeff. = 0.45) between censorship and stop word usage with a 3-week lag.

\begin{figure}[t]
\centering
\includegraphics[width=\columnwidth]{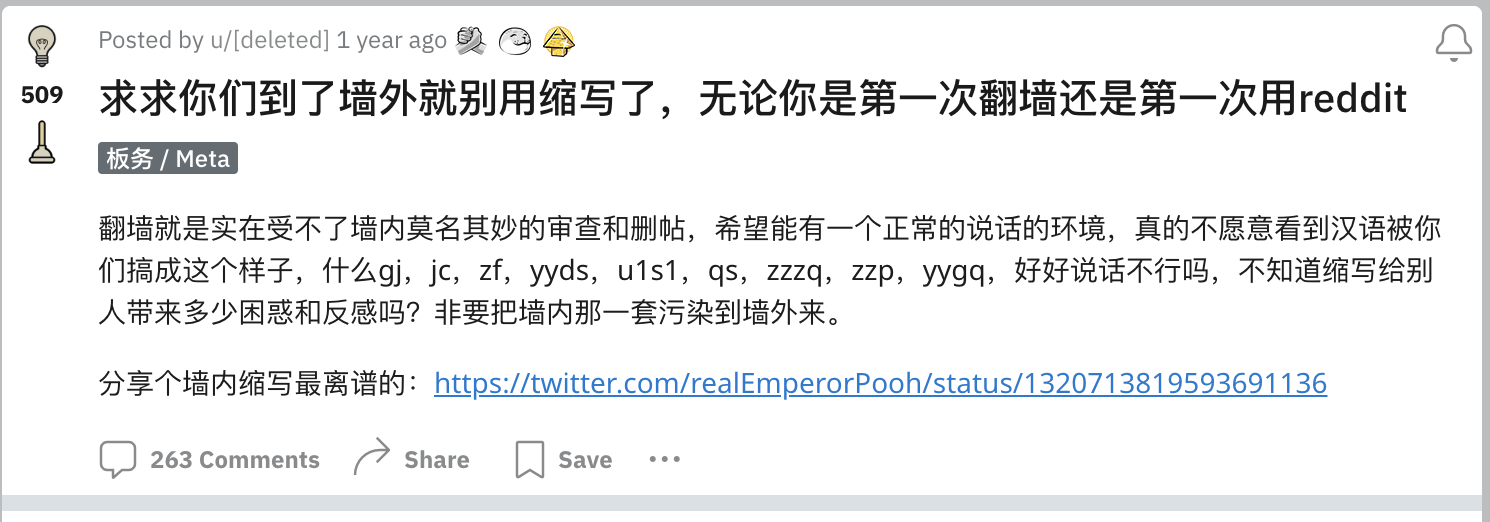} 
\caption{Translation: \textit{Please stop using abbreviations here whether or not this is your first time using Reddit! You joined Reddit with VPN because you can't stand the inexplicable censorship and deletion of posts in China. I hope to have a normal speaking environment. I really don't want to see the Chinese language become like this. What is gj, jc, zf, yyds, u1s1, qs, zzzq, zzp, yygq? can’t you speak like a human? Don’t you know how much confusion those abbreviations bring to others?}
Original link: $shorturl.at/eUV79$ }
\label{fig9}
\end{figure}

One plausible explanation for the correlation between censorship and stop word usage is a survival mechanism adopted by users to escape censorship. Users may alter their linguistic style by using less machine-detectable words to avoid censorship. However, this explanation does not seem to align with Reddit since it is not a Chinese social media platform. Interestingly, one of the labeled submissions in Figure \ref{fig1} shows a subscriber's complaint about the usage of abbreviations in the subreddit community (see Figure \ref{fig9} for translation). The complaint provides examples of sarcasm on Twitter regarding the same issue. Such collective attention to abbreviations suggests that this habit could be a migration from typical practices on Chinese social media. We are unsure if the behavior of using a lot of stop words/abbreviations is a retention of linguistic habits from Chinese social media or an imitation of the linguistic style from a censorship-heavy environment. Therefore, further research is encouraged to investigate the correlations between censorship and linguistic adaptation.

\section{Discussion and Potential}
This study opens up avenues for further research in multiple dimensions, ranging from linguistics and behavioral science to politics and natural language processing.


The results from the CDT dataset demonstrate how a topic can recur or resurface when a new, related topic emerges. \citet{cakl:16} summarized characteristics of recurring behavior of memes on Facebook, such as a moderate number of shares during the first occurrence, moderate demographic diversity, and the existence of multiple copies. While they emphasized that multiple copies act as a catalyst, the copies of the censored articles can only be considered as the ideas and topics conveyed by those articles. Our results suggest that new relevant topics stimulate the recurrence of old topics. We encourage further research to explore whether the recurrence of a series of relevant topics acts as a catalyst for social protest.

The examination of the relationships among censorship activity, comments, and subscriptions reveals correlations between censorship and comments, and between censorship and subscriptions. Our next step is to collect comments from other Chinese-speaking Subreddits to identify similarities and differences in relation to r/china\_irl. In our future work, we aim to expand our datasets to better assist researchers in studying communities and examining collective action in protests.

A limitation of this study is that the CDT dataset does not capture a complete record of all the articles censored in China. The correlations among the investigated variables largely depend on the collective archival behavior on the Wayback Machine, the database CDT collects from, which in turn is also a reflection of collective attention. Our results have shown that the increase in comments acts as a stable predictor of a coming increase in censorship. Our next step is to identify more predictors and detect and save articles with a high risk of being censored to build a more comprehensive database than CDT.

In most sentiment analysis tools, sentences are tokenized and stop words are removed to optimize the accuracy of the polarity score. In a censored environment, the extensive use of short phrases and stop words urges us to understand and explore the meanings and functions of stop words. Are these Chinese characters truly stop words? Could an NLP model for phonetic languages help us better understand the content and construction of these Reddit comments in Chinese? Assuming that the correlation between censorship and stop words usage is true, we encourage further research to investigate whether the same pattern applies to other forms of communication within autocratic environments: How does one change their linguistic habits to adapt to censorship?

Ultimately, this paper offers starting points for reflection about freedom of information, protest, and the pandemic. In China, the zero-COVID-19 policy reduced the death rate during the pandemic but triggered a series of protests. Online communities provided a virtual environment for propagating people's ideas or concerns and documenting interactions corresponding to individual reactions and behaviors in response to censorship during the pandemic. Censorship has affected the free flow of these ideas. Collective actions also exacerbated the dissemination of misinformation, such as "COVID-19 is just a big flu," across multiple scales \cite{nogara2022disinformation,pierri2023one}.


\section{Conclusions}

In the past, numerous studies have focused on the relationship between protests and COVID-19 regulations \cite{pnp:21}, or protests and censorship. Our paper presents empirical evidence to demonstrate the intricate connections between COVID-19 regulations, censorship, and protest.

The analysis of censored topic occurrences in subreddit discourses reveals several intriguing findings. First, highly sensitive topics concerning government and popular topics in r/china\_irl generate interest in similar social incidents. Second, similar topics from both datasets recur throughout 2022. Third, the censored articles and the content driving discussion in r/china\_irl slightly differ despite addressing the same issues. Censored articles tend to focus on specific social issues and emphasize consequences, while Reddit users concentrate more on seeking solutions or engaging in protest behavior to respond to these social issues.

By examining censorship activities, user subscriptions, and commenting behaviors, we identified correlations and interplays of collective attention and actions. The part of speech analysis further uncovers potential linguistic adaptations to avoid censorship.


By revealing the silenced voices through censored articles and investigating conversations in an environment with a higher degree of freedom, a series of discoveries establish possible correlations between censored articles and public agenda-setting. Further analyses are encouraged to uncover universal patterns, as such phenomena can help us understand potential survival mechanisms of marginalized groups within censored information ecosystems.

\paragraph{Acknowledgements.} Work supported in part by DARPA (contract \#HR001121C0169).

\balance
\bibliography{paper}
\bibliographystyle{aaai22}
\end{document}